\documentclass[twocolumn]{elsart}
\usepackage{graphicx,amssymb,natbib}
\def\aa{{\em A\&A}\ }

\def\aj{{\em AJ}\ }

\def\apj{{\em ApJ}\ }

\def\mnras{{\em MNRAS}\ }
\def\nat{{\em Nature}\ }

\def\lsim{\mathrel{\rlap{\lower 4pt \hbox{\hskip 1pt $\sim$}}\raise 1pt
\hbox {$<$}}} 
\def\gsim{\mathrel{\rlap{\lower 4pt \hbox{\hskip 1pt $\sim$}}\raise 1pt
\hbox {$>$}}}
\newcommand{\ms}{$M_\odot$}
\newcommand{\eg}{e.g., }

\newcommand{\Msun}{M_{\odot}}
\newcommand{\Mms}{M_{\rm MS}}
\newcommand{\kms}{km~s$^{-1}$}

\newcommand{\OI}{O~{\sc i}}
\newcommand{\SiII}{Si~{\sc ii}}
\newcommand{\CaII}{Ca~{\sc ii}}
\newcommand{\FeII}{Fe~{\sc ii}}
\newcommand{\FeIII}{Fe~{\sc iii}}
\newcommand{\TiII}{Ti~{\sc ii}}
\newcommand{\Fefs}{$^{56}$Fe}
\newcommand{\Cofs}{$^{56}$Co}
\newcommand{\Nifs}{$^{56}$Ni}
\newcommand{\Mej}{M_{\rm ej}}
\newcommand{\KE}{E_{\rm K}}

\begin{document}
\runauthor{Nomoto, Tanaka, Tominaga, Maeda, and Mazzali}
\begin{frontmatter}
\title{Hypernovae and their Gamma-Ray Bursts Connection}
\author[tokyo]{Ken'ichi Nomoto}
\author[tokyo]{Masaomi Tanaka}
\author[tokyo]{Nozomu Tominaga}
\author[MPA,tokyo2]{Keiichi Maeda}
\author[MPA,Trieste,tokyo]{Paolo A. Mazzali}

\address[tokyo]{Department of Astronomy, University of Tokyo,
  Bunkyo-ku, Tokyo 113-0033, Japan}
\address[MPA]{Max-Planck-Institut f\"ur Astrophysik,
Karl-Schwarzschild-Stra{\ss}e 1, 85741 Garching, Germany}
\address[tokyo2]{Department of Earth Science and Astronomy,
College of Arts and Science, University of Tokyo, Meguro-ku, Tokyo
153-8902, Japan}
\address[Trieste]{Istituto Nazionale di Astrofisica-OATs, Via Tiepolo
  11, I-34131 Trieste, Italy}

\begin{abstract}

The connection between long Gamma Ray Bursts (GRBs) and Supernovae
(SNe), have been established through the well observed cases of
GRB980425/SN\,1998bw, GRB030329/SN\,2003dh and GRB031203/SN\,2003lw.
These events can be explained as the prompt collapse to a black hole
(BH) of the core of a massive star ($M \sim 40 \Msun$) that had lost
its outer hydrogen and helium envelopes. All these SNe exhibited
strong oxygen lines, and their
energies were much larger than those of typical SNe, thus these SNe
are called Hypernovae (HNe).  The case of SN\,2006aj/GRB060218 appears
different: the GRB was weak and soft (an X-Ray Flash, XRF); the SN is
dimmer and has very weak oxygen lines.  The explosion energy of
SN\,2006aj was smaller, as was the ejected mass. In our model, the
progenitor star had a smaller mass than other GRB/SNe ($M \sim 20
\Msun$), suggesting that a neutron star (NS) rather than a black hole
was formed.  If the nascent neutron star was strongly magnetized (a
so-called magnetar) and rapidly spinning, it may launch a weak GRB or
an XRF. The final fate of 20-30 $\Msun$ stars show
interesting variety, as seen in the very peculiar Type Ib/c SN 2005bf.
This mass range corresponds to the NS to BH transition.  We also
compare the nucleosynthesis feature of HNe with the metal-poor stars
and suggest the Hypernova-First Star connection.

\end{abstract}


\begin{center}
{\small {\bf 
~ \\
To appear in New Astronomy Reviews \\
``A LIFE WITH STARS''\\
Amsterdam, 22-26 August 2005 \\

}}
\end{center}

\end{frontmatter}

\section{Supernovae, Hypernovae, and Gamma-Ray Bursts}

\begin{figure*}
\begin{center}
\includegraphics*[width=10.4cm]{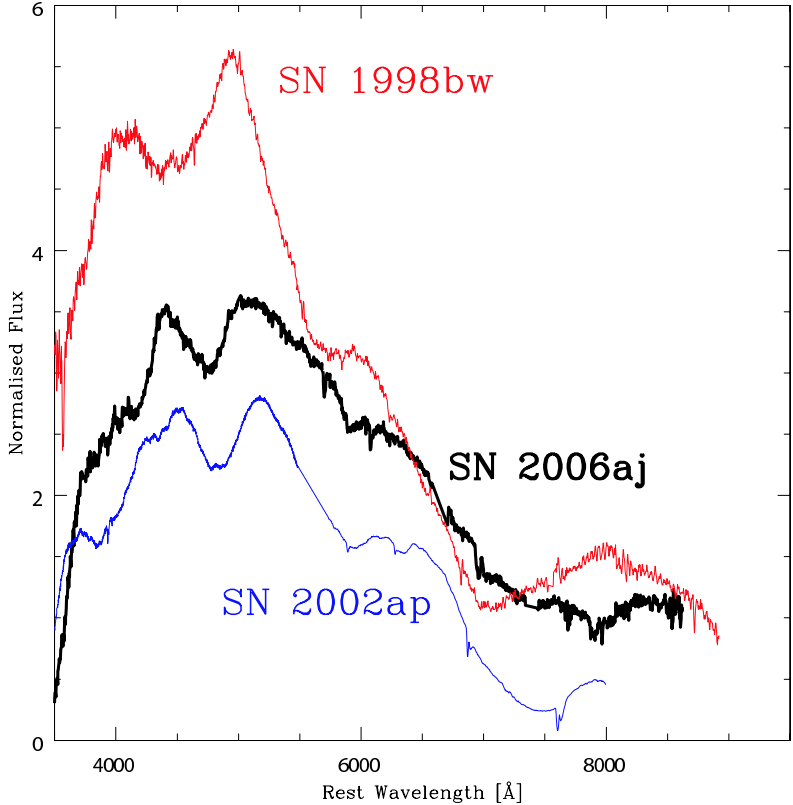}
\end{center}
\caption{
The spectra of 3 Hypernovae a few days before maximum.
SN~1998bw/GRB~980425 represents the GRB-SNe \cite{iwa98}. SN~2002ap is
a non-GRB Hypernova.  SN~2006aj is associated with XRF~060218,
being similar to SN~2002ap.
}
\label{figspcomp}
\end{figure*}

Massive stars in the range of 8 to $\sim$ 130$M_\odot$ undergo
core-collapse at the end of their evolution and become Type II and
Ib/c supernovae unless the entire star collapses into a black hole
with no mass ejection.  These Type II and Ib/c supernovae (as well as
Type Ia supernovae) release large explosion energies and eject
explosive nucleosynthesis materials, thus having strong dynamical,
thermal, and chemical influences on the evolution of interstellar
matter and galaxies.  Therefore, the explosion energies of
core-collapse supernovae are fundamentally important quantities, and
an estimate of $E \sim 1\times 10^{51}$ ergs has often been used in
calculating nucleosynthesis and the impact on the interstellar medium.
(In the present paper, we use the explosion energy $E$ for the final
{\sl isotropic} kinetic energy of explosion.)  A good example is
SN1987A in the Large Magellanic Cloud, whose energy is estimated to be
$E = (1.0$ - 1.5) $\times$ $10^{51}$ ergs from its early light curve
\citep{nomoto1994b}.

One of the most interesting recent developments in the study of
supernovae (SNe) is the discovery of the connection between
long-duration Gamma Ray Bursts (GRBs) and a particular subtype of
core-collapse SNe (Type Ic) as has been clearly established from GRB
980425/SN 1998bw \cite{gal98}, GRB 030329/SN 2003dh \cite{sta03,
hjo03}, and GRB 031203/SN 2003lw \cite{mal04}.  These GRB-SNe have
similar properties; they are all Hypernovae, i.e., very energetic
supernovae, whose {\sl isotropic} kinetic energy (KE) exceeds
$10^{52}$\,erg, about 10 times the KE of normal core-collapse SNe
(hereafter $E_{51} = E/10^{51}$\,erg).

Figure 1 shows the spectrum of SN~1998bw/GRB~980425 a few days before
maximum.  The spectrum of SN~1998bw has very broad lines, indicative
of the large $E$. The strongest absorptions are \TiII-\FeII\
(shortwards of $\sim 4000$\AA, \FeII-\FeIII\ (near 4500\AA), \SiII\
(near 5700\AA), \OI-\CaII\ (between 7000 and 8000 \AA). From the
synthetic spectra and light curves, it was interpreted as the
explosion of a massive star, with $E \sim 30 \times 10^{51}$\,erg and
$\Mej \sim 10 \Msun$.\cite{iwa98} Also the very high luminosity of SN
1998bw indicates that a large amount of \Nifs\ ($\sim 0.5 \Msun$) was
synthesized in the explosion.

The other two GRB-SNe, 2003dh and 2003lw, are also characterized by
the very broad line features and the very high luminosity.  $\Mej$ and
$\KE$ are estimated from synthetic spectra and light curves and
summarized in Figure 2\cite{nak01a, mazzali2003, deng05, maz06a}.  It
is clearly seen that GRB-SNe are the explosions of massive progenitor
stars ($M \sim 35 - 50 \Msun$), have large explosion kinetic energies
($E \sim 3 - 5 \times 10^{52}$\,erg), synthesized large amounts of
\Nifs\ ($\sim 0.3 - 0.5 \Msun$), thus forming the ``Hypernova Branch''
in Figure 2.

Other ``non-GRB Hypernovae'', such as SN~1997ef \cite{iwa00, maz00},
SN~2002ap \cite{maz02}, and SN 2003jd \cite{maz05}, have been
observed.  These HNe show spectral features similar to those of the
GRB-SNe but are not known to have been accompanied by a GRB.  The
estimated $\Mej$ and $E$, obtained from synthetic light curves and
spectra, show that there is a tendency for non-GRB HNe to have smaller
$\Mej $, $E$, and lower luminosities as summarized in Figure 2.  For
example, SN~2002ap has similar spectral features, but narrower and
redder (Fig. 1), which was modeled as a smaller energy explosion, with
$E \sim 4 \times 10^{51}$\,erg and $\Mej \sim 3 \Msun$\cite{maz02}.

Recently X-Ray Flash (XRF) 060218 has been found to be connected to SN
Ic 2006aj \cite{campana2006, pian2006}.  SN~2006aj is very similar to
SN~2002ap,
being a less energetic ($E_{51} \sim 2$).  However, it differs in the
lack of the \OI-\CaII\ absorption near 7500\AA.  
The progenitor is
estimated to be $\sim 20 M_\odot$, thus being suggested to be a
``neutron star-making SN'' \cite{mazzali2006b}
(see \S 2 for details).

\begin{figure*}
\begin{center}
\includegraphics*[width=10.4cm]{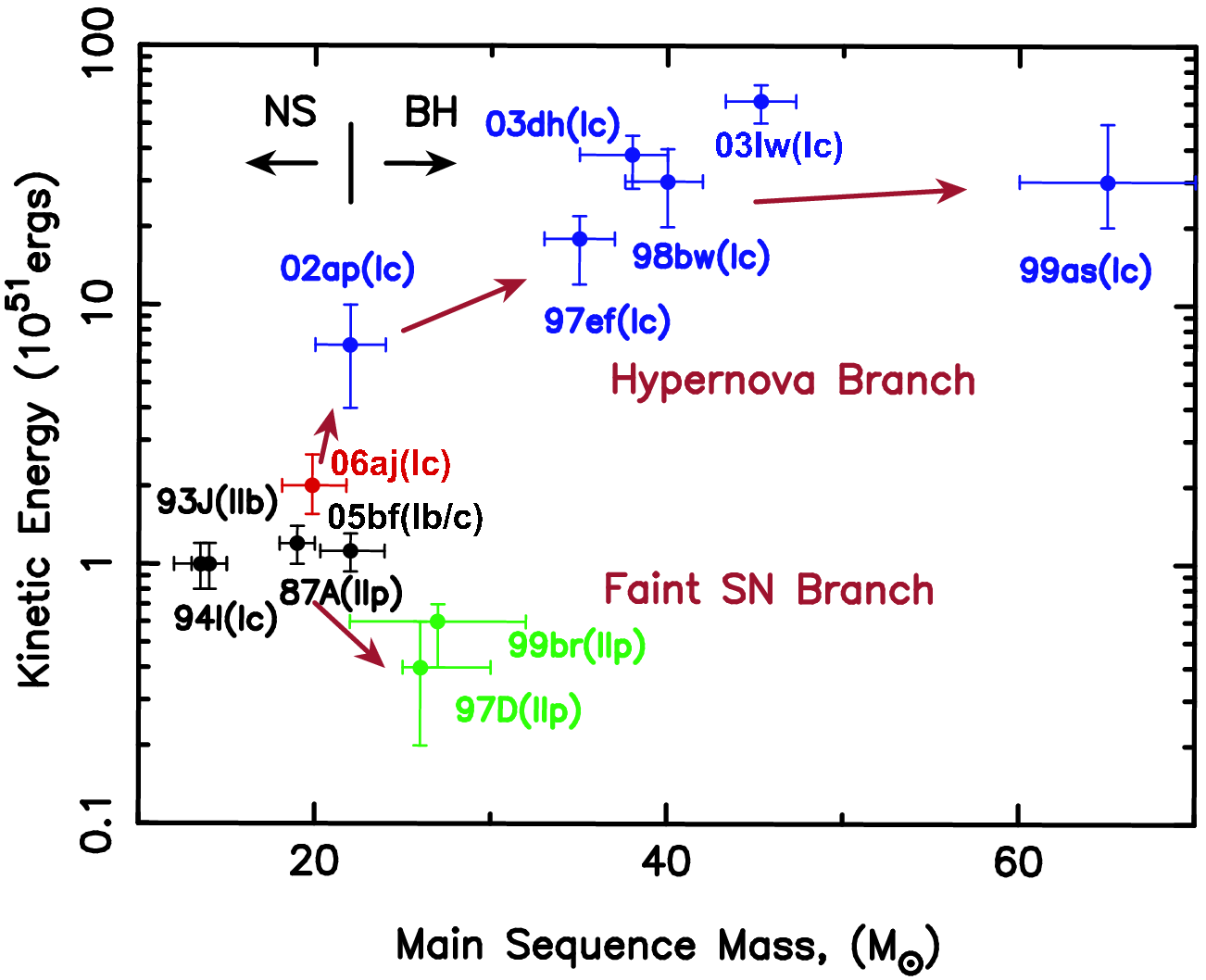}\\
\includegraphics*[width=10.4cm]{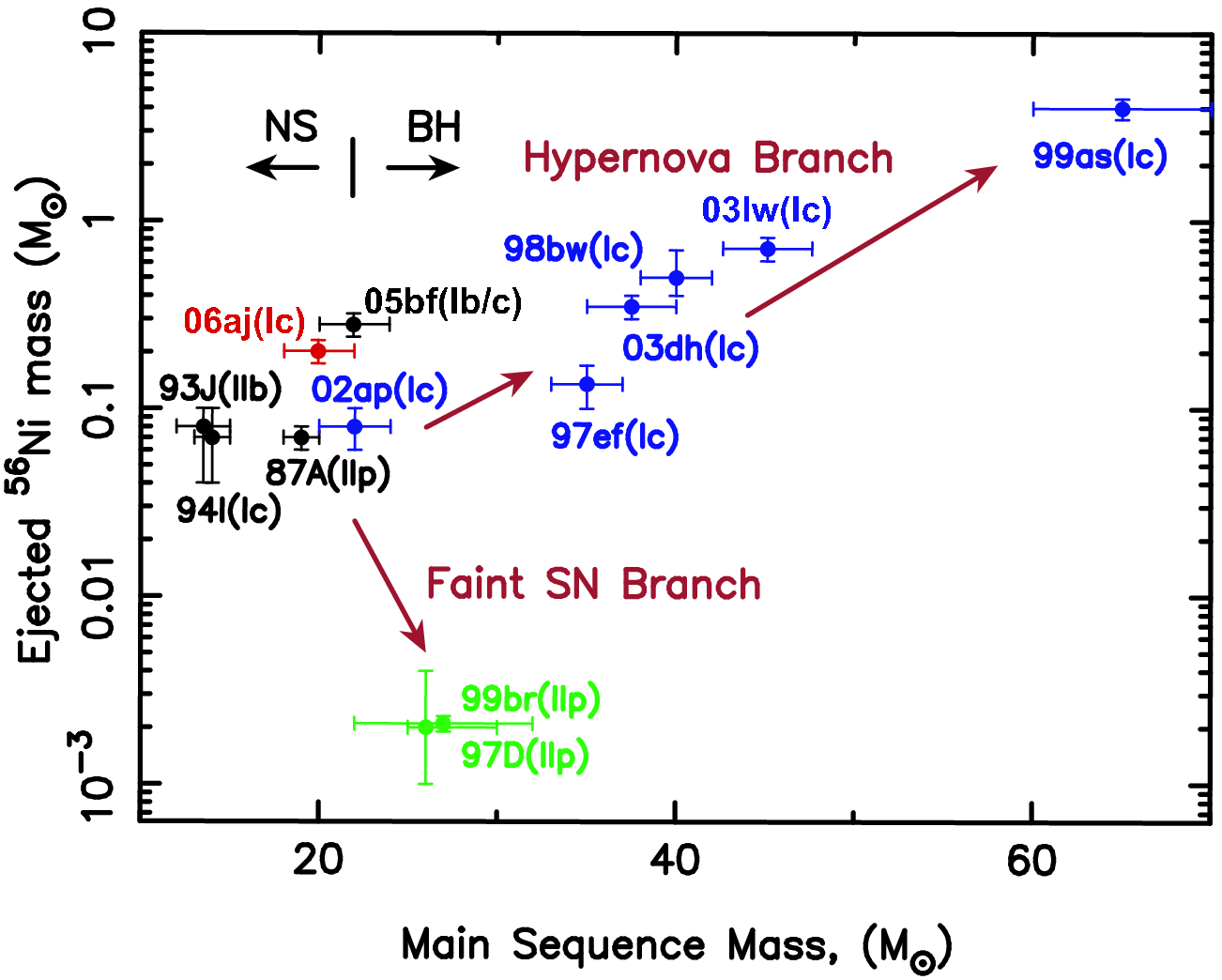}
\end{center}
\caption{
The kinetic explosion energy $E$ and the ejected $^{56}$Ni mass as a
function of the main sequence mass $M$ of the progenitors for several
supernovae/hypernovae.}
\label{figME}
\end{figure*}

Hypernovae are also characterized by asphericity from the observations
of polarization and emission line features \cite[e.g.,][]{wang2003,
kaw02, mae02}.  The explosion energy of the aspherical
models for hypernovae tends to be smaller than the spherical models by
a factor of 2 - 3, but still being as high as $E_{51} \gsim 10$
\cite{mae06LC}.

In contrast to HNe, SNe II 1997D and 1999br were very faint SNe with
very low KE \cite{turatto1998, hamuy2003, zampieri2003}.  In the
diagram that shows $E$ and the mass of $^{56}$Ni ejected $M(^{56}$Ni)
as a function of the main-sequence mass $M_{\rm ms}$ of the progenitor
star (Figure~\ref{figME}), therefore, we propose that SNe from stars
with $M_{\rm ms} \gsim 20-25 M_\odot$ have different $E$ and
$M(^{56}$Ni), with a bright, energetic ``hypernova branch'' at one
extreme and a faint, low-energy SN branch at the other
\cite{nomoto2003}.  For the faint SNe, the explosion energy was so
small that most $^{56}$Ni fell back onto the compact remnant
\cite[e.g.,][]{sollerman1998}.  Thus the faint SN branch may become a
``failed'' SN branch at larger $M_{\rm ms}$.  Between the two
branches, there may be a variety of SNe \cite{hamuy2003,
tominaga2005}.

This trend might be interpreted as follows.  Stars more massive than
$\sim$ 25 $M_\odot$ form a black hole at the end of their evolution.
Stars with non-rotating black holes are likely to collapse ``quietly''
ejecting a small amount of heavy elements (Faint supernovae).  In
contrast, stars with rotating black holes are likely to give rise to
Hypernovae.  The hypernova progenitors might form the rapidly rotating
cores by spiraling-in of a companion star in a binary system.

Here we focus on our findings that SN Ic 2006aj (\S2) and SN Ib 2005bf
(\S3) are very different SNe from previously known SNe/HNe.  These
properties might be due to their progenitor masses, which indicate
that these SNe correspond to the border from the NS and BH formation.

As a related topic, it is of vital importance to identify the first
generation stars in the Universe, i.e., totally metal-free, Pop III
stars.  We examine possible GRB/Hypernova - First Star connection
through nucleosnthesis approach \cite{nomoto2006}.  We summarize
nucleosynthesis features in hypernovae, which must show some important
differences from normal supernova explosions (\S4).  This might be
related to the unpredicted abundance patterns observed in the
extremely metal-poor halo stars.  This is one of the important
challenges of the current astronomy~\cite{weiss2000,abel2002}.

\section{SN~2006aj}

GRB060218 is located in a galaxy only $\sim 140\,$Mpc away and it is
the second closest event as ever.  The GRB was weak
\cite{campana2006}, as is often the case for nearby ones \cite{sod04},
and was classified as X-Ray Flash (XRF) because of its soft spectrum.
As the GRB was not followed by a bright afterglow, the presence of a
SN 2006aj was soon confirmed \cite{pian2006,mod06}.  Here we summarize 
the properties of SN 2006aj by comparing with other SNe~Ic.

\begin{figure*}
\begin{center}
\includegraphics*[width=12.4cm]{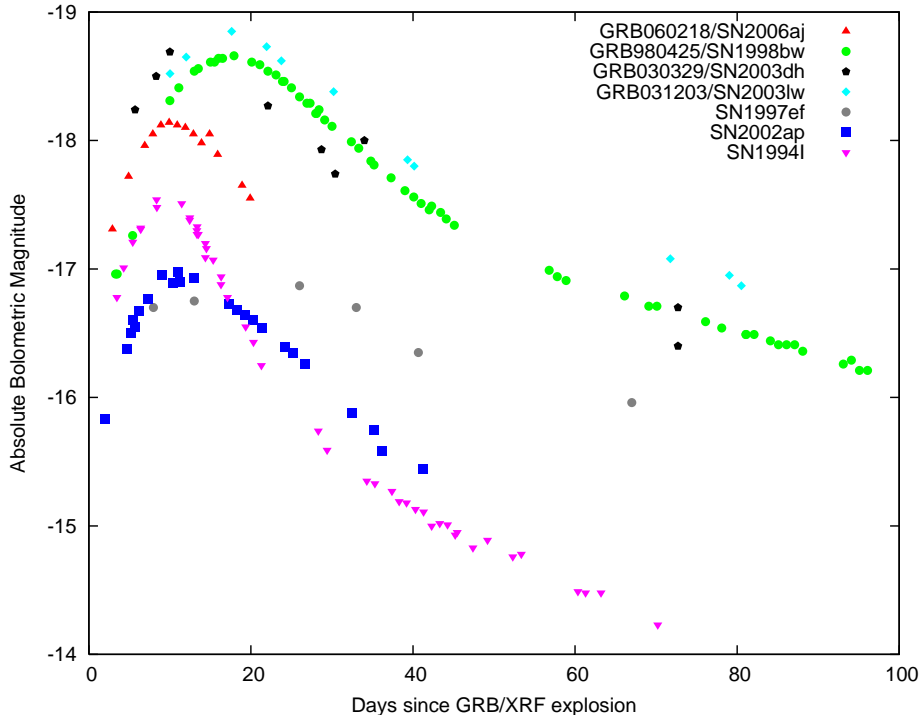}
\end{center}
\caption{The bolometric LC of SN 2006aj compared with 
other GRB/SN (SN 1998bw, SN 2003dh) and non-GRB/SN (SN 1997ef).}
\label{fig06ajLC}
\end{figure*}

SN~2006aj has several features that make it unique.  It is less bright
than the other GRB/SNe (Figure \ref{fig06ajLC}).  Its rapid
photometric evolution is very similar to that of a dimmer, non-GRB SN
2002ap \cite{maz02}, but it is somewhat faster.  Although its spectrum
is characterized by broader absorption lines as in SN 1998bw and other
GRB/SN, they are not as broad as those of SN~1998bw, and again it is
much more similar to that of SN~2002ap (Figure \ref{fig06ajsp}).  The
most interesting property of SN~2006aj is surprisingly weak oxygen
lines, much weaker than in Type Ic SNe.

\subsection{Spectroscopic Models}

In order to quantify its properties, we modeled the spectra of
SN~2006aj with a radiation transfer code as in \cite{mazzali2000}.  
We first
employed the same explosion model applied for SN 2002ap \cite{maz02}.
The model has an ejected mass $\Mej$ $\sim 3 \Msun$ and a kinetic
energy $E \sim 4 \times 10^{51}$\,erg.  This explosion model gives
reasonable fits to the spectra.
 
However, in order to improve the match, we had to reduce the masses of
both oxygen and calcium substantially and reduce $\Mej$ accordingly
(Figure \ref{figspfit}).  As a result, we derive for SN~2006aj $\Mej
\sim 2 \Msun$ and $E \sim 2 \times 10^{51}$\,erg.  Lack of oxygen in
the spectra does not necessarily mean absence of oxygen in the ejecta.
Our model contains $\sim 1.3 \Msun$ of O, and oxygen is still the
dominant element.

\begin{figure*}
\begin{center}
\includegraphics*[width=9.4cm]{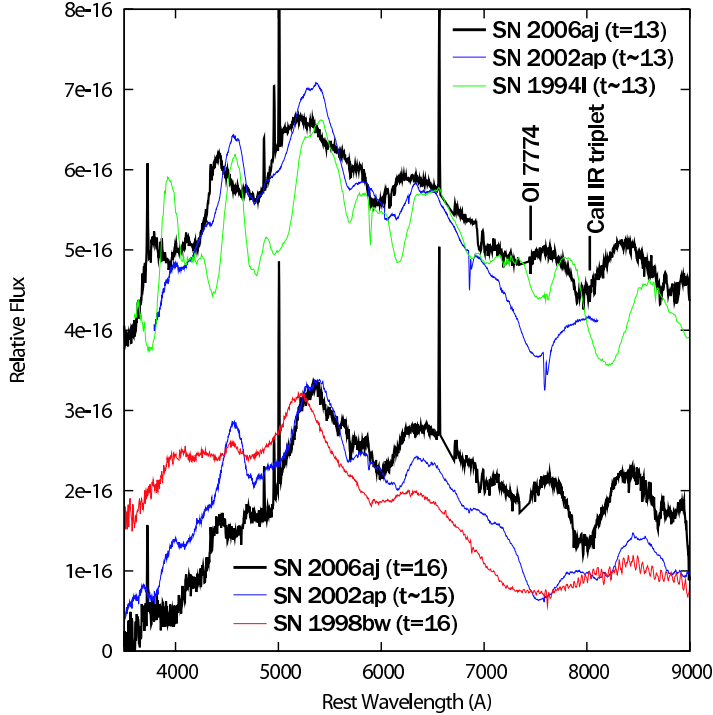}
\end{center}
\caption{({\it top}) The optical spectrum of SN 2006aj 
at 13 days since the GRB/XRF (bold line) compared with the spectrum 
of SN 2002ap (thin gray line) and SN 1994I (dotted line) 
at $\sim$ 13 days since the explosion.
({\it bottom}) The optical spectrum of SN 2006aj 
at 16 days since the GRB/XRF (bold line) compared with the spectrum 
of SN 2002ap at $\sim$ 15 days since the explosion (thin gray line) 
and SN 1998bw at 16 days since GRB980425 (dotted line). 
}
\label{fig06ajsp}
\end{figure*}

\begin{figure*}
\begin{center}
\includegraphics*[width=9.4cm]{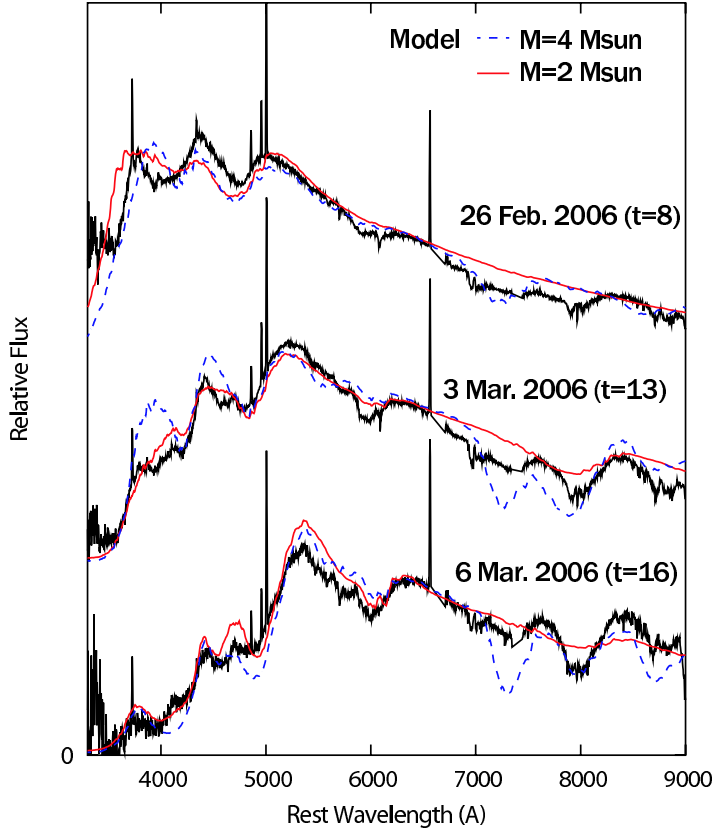}
\end{center}
\caption{Synthetic spectra with a explosion model with 
$(\Mej, E)$ = $(2.0 \Msun , 2.0 \times 10^{51}$ergs) [gray solid lines]
and $(4.0 \Msun , 9.0 \times 10^{51}$ergs) [black dashed lines]
compared with the observed spectra of SN 2006aj (solid lines).}
\label{figspfit}
\end{figure*}

The strength of the OI$\lambda$7774 line, which is the strongest
oxygen line in optical wavelength, is sensitive to the temperature in
the ejecta.  Since the fraction of OI is larger in the lower
temperature ejecta (although OII is still the dominant ionization
state), the normal SNe Ib/c always show the strong OI absorption (see
SN 1994I in Fig \ref{fig06ajsp}) irrespective of the ejecta mass.

In more luminous SNe like GRB-SNe and SN 2006aj, the OI fraction tends
to be smaller.  However, if the ejecta are very massive, \eg $\sim 10
\Msun$, the mass of OI is large enough to make the strong absorption
(see SN 1998bw in Fig \ref{fig06ajsp}).  In the case of SN 2006aj, the
temperature is larger than in normal SNe Ib/c.  Therefore, the weak OI
line indicates that the ejecta mass is not as massive as SN 1998bw,
which supports our conclusion.

\subsection{Light Curve Models}

The spectroscopic results are confirmed through the light curve
modeling.  The timescale of the LC around maximum brightness reflects
the timescale for optical photons to diffuse \cite{arn82}.  For the
more massive ejecta and the smaller kinetic energy, the LC peaks later
and the LC width becomes broader becaause it is more difficult for
photons to escape.

\begin{figure*}
\begin{center}
\includegraphics*[width=9.4cm]{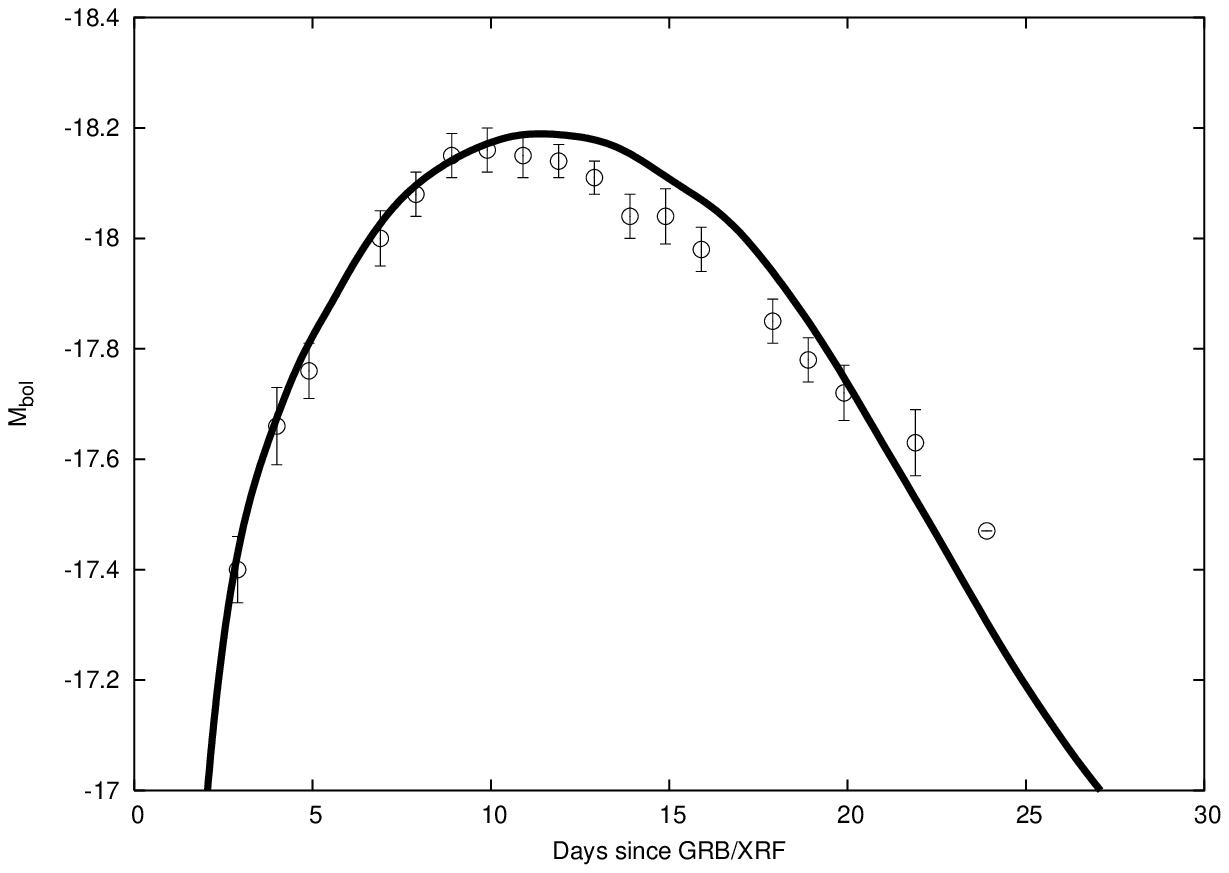}
\end{center}
\caption{The synthetic LC computed for the model with
$(\Mej, E)$ = $(2.0 \Msun , 2.0 \times 10^{51}$ergs).}
\label{figLCfit}
\end{figure*}

We synthesize the theoretical light curve with the 1-dimensional
density and chemical abundance structure that we find in the above
spectroscopic analysis.  We then compare it with the optical-infrared
bolometric light curve of SN~2006aj.  The best match to the rapidly
rising light curve is achieved with a total \Nifs\ mass of $0.21
\Msun$ in which $0.02 \Msun$ is located above 20,000\kms (Figure
\ref{figLCfit}).  The high-velocity \Nifs\ is responsible for the fast
rise of the light curve, because photons created can escape more
easily.

In the model, the mass fraction of \Nifs\ in the high velocity region
is as large as $\sim$ 35\%, which is unlikely to be attained in a
spherically symmetric explosion.  In a realistic asymmetric explosion,
the high-velocity $^{56}$Ni could abundantly be produced along the
direction of the GRB jets \cite{mae02,mae03}.

\subsection{The Progenitor and Implications for XRF}

The properties of the SN~2006aj (smaller energy, smaller ejected mass)
suggest that SN~2006aj is not the same type of event as the other
GRB-SNe known thus far.  One possibility is that the initial mass of
the progenitor star is much smaller than the other GRB-SNe, so that
the collapse/explosion generated less energy.  If the zero-age main
sequence mass is $\sim 20 - 25 \Msun$, for example, the star would be
at the boundary between collapse to a black hole or to a neutron star.
In this mass range, there are indications of a spread in both $E$ and
the mass of \Nifs\ synthesized \cite{hamuy2003}.  The fact that a
relatively large amount of \Nifs\ is required in SN 2006aj possibly
suggests that the star collapsed only to a neutron star because more
core material would be available to synthesize \Nifs\ in the case.

Although the kinetic energy of $\sim 2 \times 10^{51}$ erg is larger
than the canonical value ($1 \times 10^{51}$ erg, \cite{nom94}) in the
mass range of $M_{\rm ms} \sim 20 - 25 \Msun$, such an energy may be
easily attainable.  Additionally, magnetar-type activity may have been
present, increasing the explosion energy\cite{tho04}.  It is
conceivable that in this weaker explosion than typical GRB-SNe, the
fraction of energy channeled to relativistic ejecta is smaller, giving
rise to an XRF rather than a classical GRB.  

Another case of a SN associated with an XRF has been
reported (XRF030723)\cite{fyn04}.  The putative SN, although its spectrum was not
observed, was best consistent with the properties of
SN~2002ap \cite{tom05}.  This may suggest that XRFs are associated with
less massive progenitor stars than those of canonical GRBs, and that
the two groups may be differentiated by the formation of a neutron
star \cite{nak98} or a BH.  Still, the progenitor star must have been
thoroughly stripped of its H and He envelopes, which is a general
property of all GRB-SNe and probably a requirement for the emission of
a high energy transient.  These facts may indicate that the progenitor
is in a binary system.

If magnetars are related to the explosion mechanism, some short
$\gamma$-ray repeaters energized by a magnetar \cite{tho95,tho04} may
be remnants of GRB060218-like events.  Magnetars could generate a GRB
at two distinct times.  As they are born, when they have a very large
spin rate ($\sim 1$ ms), an XRF (or a soft GRB) is produced as in
SN\,2006aj/GRB060218. Later (more than 1,000 yrs), when their spin
rate is much slower, they could produce short-hard GRBs \cite{hur05}.

Stars of mass $20-25 \Msun$ are much more common than stars of $35-50
\Msun$, and so it is highly likely that events such as GRB060218 are
much more common in nature than the highly energetic GRBs.  They are,
however, much more difficult to detect because they have a low
$\gamma$-ray flux.  The discovery of GRB060218/SN~2006aj suggests that
there may be a wide range of properties of both the SN and the GRB in
particular in this mass range.  The continuing study of these
intriguing events will be exciting and rewarding.

\section{SN~2005bf} 
\label{sec:sn05bf}

Peculiar and inhomogeneous natures of supernovae originating from
$M_{\rm ms} \sim 20 - 30\Msun$, i.e., boundary between a neutron star
formation and a black hole formation, are further highlighted by SN
2005bf.  Unlike SN 2006aj/GRB060218, it did not show a high energy
transient counter part. However, it did show very unique photometric
and spectroscopic behavior.  Based on the calculated light curve and
the spectra, we believe that SN2005bf fits into the scheme suggested
by \cite{nom95} which places core-collapse SN in a sequence
(IIP-IIL-IIb-Ib-Ic) of increasing mass loss from the progenitor star,
and that it originated in collapse of a star with $M_{\rm ms} \sim
25\Msun$.

SN 2005bf was discovered by \cite{mon05,moo05} on April 6, 2005 (UT)
in the spiral arms of the SBb galaxy MCG +00-27-5.  It was initially
classified as a Type Ic SN (SN Ic) \cite{mor05,mod05a}.  As time went
by, He lines were increasingly developed.  Then it was classified as
Type Ib \cite{wan05,mod05b}.  Even stranger, the light curve is very
different from any known SN \cite{ham05}: a fairly rapid rise to a
first peak was followed by a period of stalling or slow decline and by
a new rise to a later, brighter peak at $\sim 40$ days after explosion
(Fig.~\ref{fig05bfLC}).  The brightness ($M_{\rm bol}\sim -18$ mag) at
the relatively slow peak date suggests that a large amount of \Nifs\
is ejected. SN 2005bf does not show the broad lines seen in
hypernovae.  These properties make SN~2005bf a very interesting SN.

\subsection{Early Phases}
\label{sec:bf-early}

\begin{figure*}
\begin{center}
\includegraphics*[width=8.5cm]{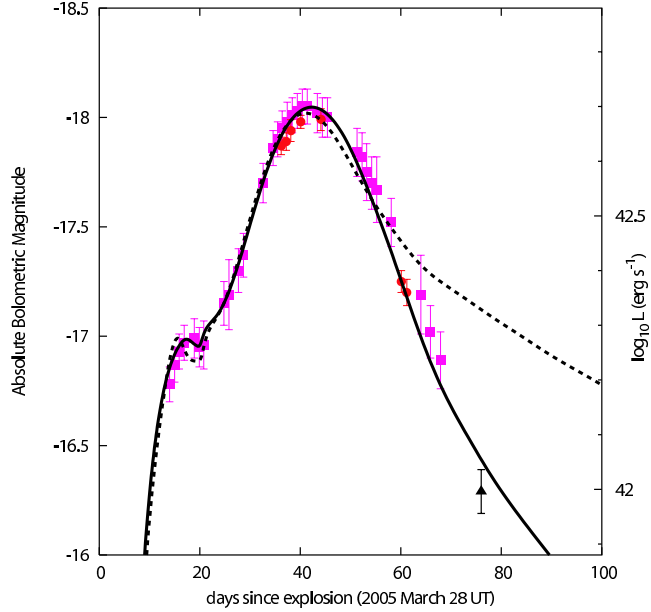}
\end{center}
\caption{The bolometric light curve of SN2005bf, constructed 
from FLWO (filled squares; 
\cite{mod05c}), HCT (filled circles; \cite{anu05}), 
and MAGNUM (filled triangle) photometry. 
Synthetic light curves are shown for normal (dashed) and reduced (solid) 
$\gamma$-ray opacities (see text).}
\label{fig05bfLC}
\end{figure*}

To derive physical quantities from observed properties, we first try
fitting the light curve covering the first 80 days \cite{tom05b}.  The
radioactive decay (\Nifs\ $\to$ \Cofs\ $\to$\Fefs) is responsible for
powering the light curve.  The theoretical LC width near peak depends
on ejected mass $\Mej$ and explosion kinetic energy $E$ as $\Mej
E^{-3}$ \cite{arn82,nom04}. The mass and distribution of \Nifs\ are
constrained by the LC brightness and shape.  In general, various
combinations of ($\Mej$, $E$) can fit the LC.  The degeneracy can be
solved by performing spectral modeling, since the line width is scaled
as $E^{1/2} M^{-1/2}$.

We compute a set of synthetic light curves for a He star model, by
varying ($\Mej$, $E$) and abundance distribution (including
distribution of the heating source $^{56}$Ni).  Parameters are
constrained by the observed bolometric light curve (LC) constructed as
in \cite{yos03} (see Fig.~\ref{fig05bfLC}).  We assumed a Galactic
reddening $E(B-V)=0.045$, a distance modulus $\mu=34.5$, and an
explosion date of $2005$ March 28 UT as inferred from the marginal
detection on 2005 March 30 UT \citep{moo05}.  The light curve for a
model with $\Mej=7\Msun$ and $E_{51} = E/10^{51}{\rm ergs}=2.1$ is
shown as a dashed line in Figure~\ref{fig05bfLC}.

The model curve yields a nice fit until the second, main peak is
reached.  However, the observed LC declines rapidly thereafter, unlike
other well-observed SNe Ib/c.  A possible solution for this is to
consider a situation in which the ejecta are more transparent to
gamma-rays than in other SNe Ib/c (usually $\kappa_\gamma = 0.025 {\rm
cm^2 g^{-1}}$ \cite{mae06}).  Using $\kappa_\gamma = 0.001 {\rm cm^2
g^{-1}}$ at $v<5,400$\,\kms\, the light curve shown in
Figure~\ref{fig05bfLC} is obtained.  The model parameters are the
following: $\Mej=7\Msun$, $E_{51} = 1.3$, $M({\rm
^{56}Ni})=0.32\Msun$.  The parameters are not so different from those
derived without accelerated gamma-ray escape, since the set of
($\Mej$, $E$) is constrained by the diffusion time scale and expansion
time scale, and $M$($^{56}$Ni) is determined from the peak
brightness. These are basically independent from the LC behavior after
the peak date.

\begin{figure*}
\begin{center}
\includegraphics*[width=9.5cm]{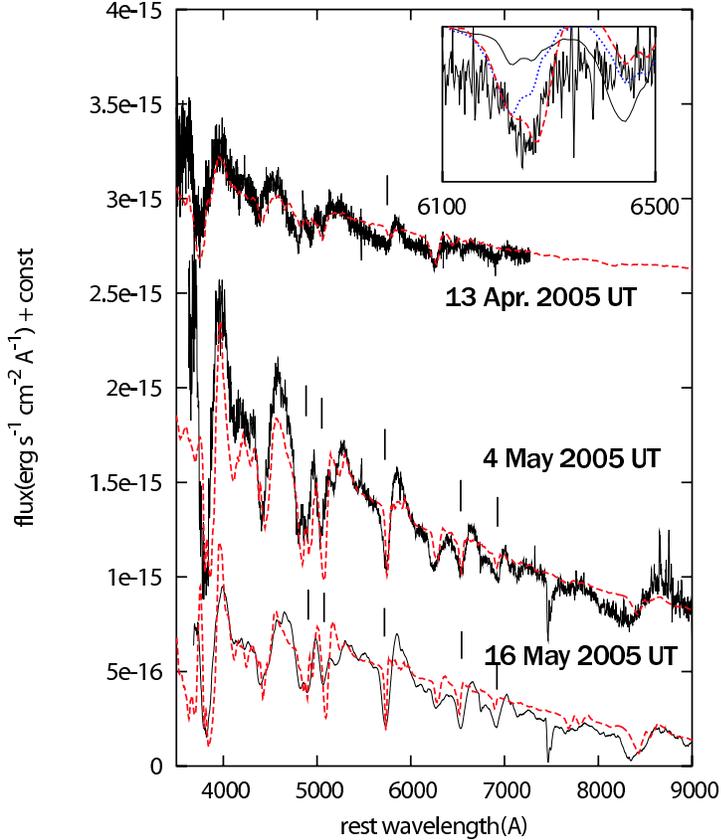}
\end{center}
\caption{
Spectra of SN~2005bf (thick lines: 2005 April 13 - FLWO,
\cite{mod05c}; May 4 - HCT, \cite{anu05}; May 16 - Subaru Telescope) 
compared to the synthetic spectra (dashed lines)
computed with the model ($\Mej/\Msun$, $E_{51}$) = (7, 1.3). The
position of He lines is shown by tick marks.  The absorptions near
4900 and 5100\AA\ are blended with Fe {\scriptsize II} lines.  The
inset shows the absorption near 6300\AA\ in the April 13th spectrum.
The model with H at $v \gsim\ 13,000$ \kms\ (dashed line) provides the
best fit. Thin and dotted lines show models with H in the whole ejecta
and no H, respectively.
}
\label{fig05bfsp}
\end{figure*}

To break the degeneracy in ($\Mej$, $E$) and select the most likely
model, synthetic spectra are computed and compared to the observed
ones in Figure \ref{fig05bfsp}.  The model with ($\Mej/\Msun$,
$E_{51}$) = (7, 1.3) provides satisfactory fits for all spectra.

At the time of the first peak (UT April 13, 16 days after explosion; \cite{mod05c}) 
SN~2005bf exhibited SN Ic features, but actually both He and H can be seen 
as weak features. 
The feature near 5700\AA\ is probably He {\scriptsize I} 5876\AA. 
A photospheric velocity of $v_{\rm ph}=6,200$\,\kms is obtained by the spectroscopic 
model. 
The model with $\Mej=7\Msun$ fits better than that with $\Mej=6\Msun$. 
The feature at 6300\AA\ is partly attributed to 
H$\alpha$ (Fig.~\ref{fig05bfsp}, inset), 
indicating presence of $\sim 0.02\Msun$ of hydrogen above $v \gsim 13,000$ \kms. 
The spectra near 
maximum brightness (UT May 4, 37 days after explosion) 
and on UT May 16th (49 days) are 
reasonably reproduced by the same ejecta model, 
with the photospheric velocity $v_{\rm ph}=4,600$ km s$^{-1}$ 
and $3,800$ km s$^{-1}$, respectively. 

According to the model, the ejecta properties are derived as follows:
$\Mej \sim 7\Msun$ and $E_{51} \sim 1.3$.  The ejecta consist of
\Nifs\ ($\sim 0.32 \Msun$) mostly near the center, He ($\sim
0.4\Msun$), intermediate mass elements (mainly O, Si, S), and a small
amount of H ($\sim 0.02\Msun$). Thus the progenitor had lost almost
all its H envelope, but retained most of the He-rich layer (a WN
star).

The He core mass at the explosion was $M_{\rm He}=\Mej+M_{\rm cut}\sim
7.5-8.5\Msun$. The progenitor was probably a WN star of main-sequence
mass $\Mms \sim25\Msun$ \citep{nom88,ume05}.  The formation of a WN
star from a star of only $\sim 25\Msun$ suggests that rotation may
have been important \citep{hir05}, although not sufficient to make SN
2005bf a hypernova.  In order to produce $\sim 0.32\Msun$ \Nifs, the
mass cut that separates the ejecta and the compact remnant should be
as deep as $M_{\rm cut}\sim 1.4\Msun$.  This suggests that the remnant
was a neutron star rather than a black-hole.  The mass range $\Mms
\sim 25\Msun$ is near the transition from neutron star (SN 2005bf) to
black hole formation (SN 2002ap; \cite{maz02}), the exact boundary
depending on rotation and mass loss.

The ejecta we derived have small $E$ relative to $\Mej$. In this way,
relatively delayed peak at $\sim 40$ days is explined.  The rapid rise
to the first peak at $\sim 20$ days requires a small amount of
$^{56}$Ni ($\sim 0.06 \Msun$) at high velocity ($v \gsim 3,900$ \kms),
while most of $^{56}$Ni is neat the center below $1,600$ km s$^{-1}$.
The high velocity component of \Nifs\ is at the bottom of the He
layer.  It may express $^{56}$Ni-rich jets or blobs that did not reach
the He layer.  Such inhomogeneous structure might result in the
enhanced $\gamma$-ray escape as is assumed to obtain a better fit to
the LC after the second peak.

\subsection{Late Phases}
\label{sec:bf-late}

At late epochs ($\sim 1$ year since the explosion), a SN enters into
nebular phases. Then a SN shows optically thin spectra dominated by
emission lines arising even from the deepest regions.  Thus the
nebular phase observations can prove the inner regions of the ejecta,
which are not seen at early phases.  Spectroscopy and photometry of SN
2005bf have been performed on 2005 December 26 (UT) and on 2006
February 6 with the 8.2 m Subaru telescope equipped with the Faint
Object Camera and Spectrograph (FOCAS; \cite{kas02}).  Detailed
analysis for the late time observation is presented elsewhere
\cite{mae06b}.

\begin{figure*}
\begin{center}
\includegraphics*[width=12.5cm]{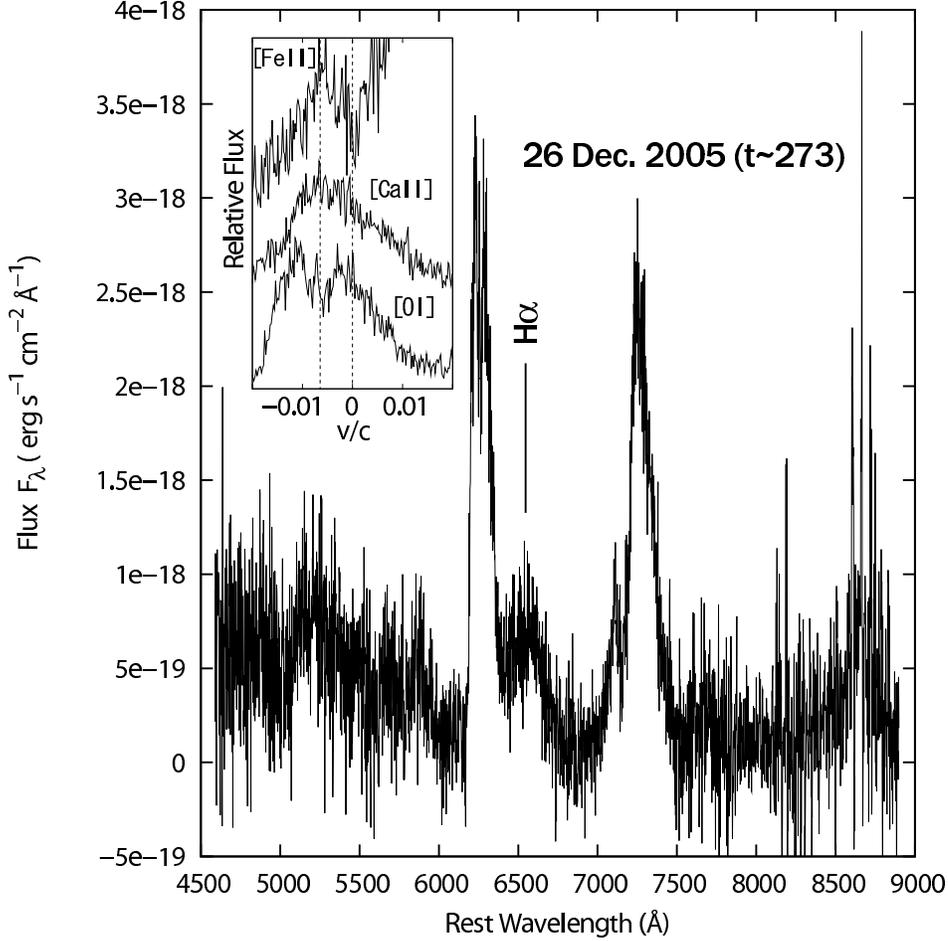}
\end{center}
\caption{A nebular Spectrum of SN 2005bf}
\label{figneb}
\end{figure*}

Figure \ref{figneb} shows the reduced spectra of SN 2005bf on 2005
December 26 ($\sim 270$days since the putative explosion date).  There
is a feature at $\sim 6,500$\AA\ with FWHM $\sim 15,000$ km s$^{-1}$.
This is most likely H$_{\alpha}$ emission.  This supports the
existence of the thin H envelope described in \S \ref{sec:bf-early}

The spectrum shows strong forbidden emission lines, in which [OI]
$\lambda\lambda$6300, 6363 and [CaII] $\lambda$7300 are the strongest.
Other weak emission features are visible at $\sim 5200$\AA\ ([FeII])
and $\sim 8700$\AA (CaII IR and [CI] $\lambda$8727).  The [OI]
$\lambda$6300/[CaII] $\lambda$7300 ratio in SN 2005bf is relatively
small, indicating the progenitor star with $M_{\rm ms} < 40\Msun$.  A
forbidden carbon line [CI] $\lambda$8727 looks strong, indicating high
C/O ratio ($\sim 35\%$) in the ejecta.  This suggests the progenitor
star with $M_{\rm ms} \lsim 20 - 25\Msun$.  From these analyses, the
nebular spectra supports the progenitor mass $M_{\rm ms} \sim
25\Msun$, as derived in the earlier phases (\S \ref{sec:bf-early}).

Profiles of these line profiles are shown in Figure
\ref{figneb}. Interestingly, all these lines show blueshift relative
to the rest wavelength ($\sim 1,500 - 2,000$ km s$^{-1}$).  A simple
and straightforward interpretation is that we see a unipolar
explosion, on average moving toward us, as is expected from the
fast-moving $^{56}$Ni seen in the early phase (\S \ref{sec:bf-early}).
Another interpretation is self-absorption of the light within the
ejecta which reduces the light from the far (therefore red) side.
These are thoroughly discussed in Maeda et al. (2007) \cite{mae06b}.

\subsection{Peculiarities}
\label{bf-pec}

The major features of SN 2005bf can be understood in the context of
the explosion of a WN star with the progenitor $M_{\rm ms} \sim 25
\Msun$.  However, there are still some questions to be answered, which
are probably related to unique natures of the progenitor, the central
remnant, and the explosion physics of stars with $M_{\rm ms} \sim 20 -
30\Msun$.

First, in early phases, He lines evolved in their strengths and
velocities in a unique way.  They become stronger (therefore look like
SN Ic first, then like SN Ib afterward) as time goes by.  Their
velocities are also increased as a function of time.  At the first
peak, these lines are well explained in LTE level populations.  Near
maximum brightness (UT May 4, 37 days after explosion) and on UT May
16th (49 days), however, level populations of He ion
should be more abundant than in LTE populations by a factor of $\sim
2.0 \times 10^{3}$ (at $v \gsim 6,500 $ \kms) and $\sim 2 \times10^6$
(at $v \gsim 7,200 $ \kms), respectively.

Non-thermal effects resulting from radioactive decay gamma-rays are
believed to be essential to populate He{\scriptsize I} levels
\citep{luc91}.  Why SN 2005bf showed the peculiar evolution of He
lines is probably related to its unique $^{56}$Ni distribution.  This
is still an open question, which will probably provide further clue to
understand the natures of SN 2005bf.

Another question was brought by the nebular phase observation.  The
R-band magnitude at 2006 December 26 ($\sim 270$ days since the
explosion) is $\sim 24.4$, corresponding the absolute magnitude $\sim
-10.2$ after correcting the distance and the reddening.  It is very
faint as compared to other SNe Ib/c, at least by 2 magnitudes (e.g.,
by 3 magnitudes fainter than SN 1998bw at a similar epoch).  If the R
magnitude is close the bolometric magnitude (which is usually a good
approximation for SN Ib/c nebulae if most of the light is emitted in
optical ranges), $^{56}$Ni required to fit the luminosity is only
$\sim 0.03 - 0.08\Msun$.  Why this is much smaller than that
reproducing the early phase peak luminosity (at $\sim 40$ days) is
still to be answered.  Several possibilities, including late time
fallback of materials onto a central remnant and a magnetar-like
activity are worth studying \cite{mae06b}.

\section{Nucleosynthesis in Hypernovae and The First Star Connection}

In core-collapse supernovae/hypernovae, stellar material undergoes
shock heating and subsequent explosive nucleosynthesis. Iron-peak
elements are produced in two distinct regions, which are characterized
by the peak temperature, $T_{\rm peak}$, of the shocked material.  For
$T_{\rm peak} > 5\times 10^9$K, material undergoes complete Si burning
whose products include Co, Zn, V, and some Cr after radioactive
decays.  For $4\times 10^9$K $<T_{\rm peak} < 5\times 10^9$K,
incomplete Si burning takes place and its decayed products include
Cr and Mn (e.g., \cite{nakamura1999}).

\begin{figure*}
\centering
\includegraphics*[width=6.5cm]{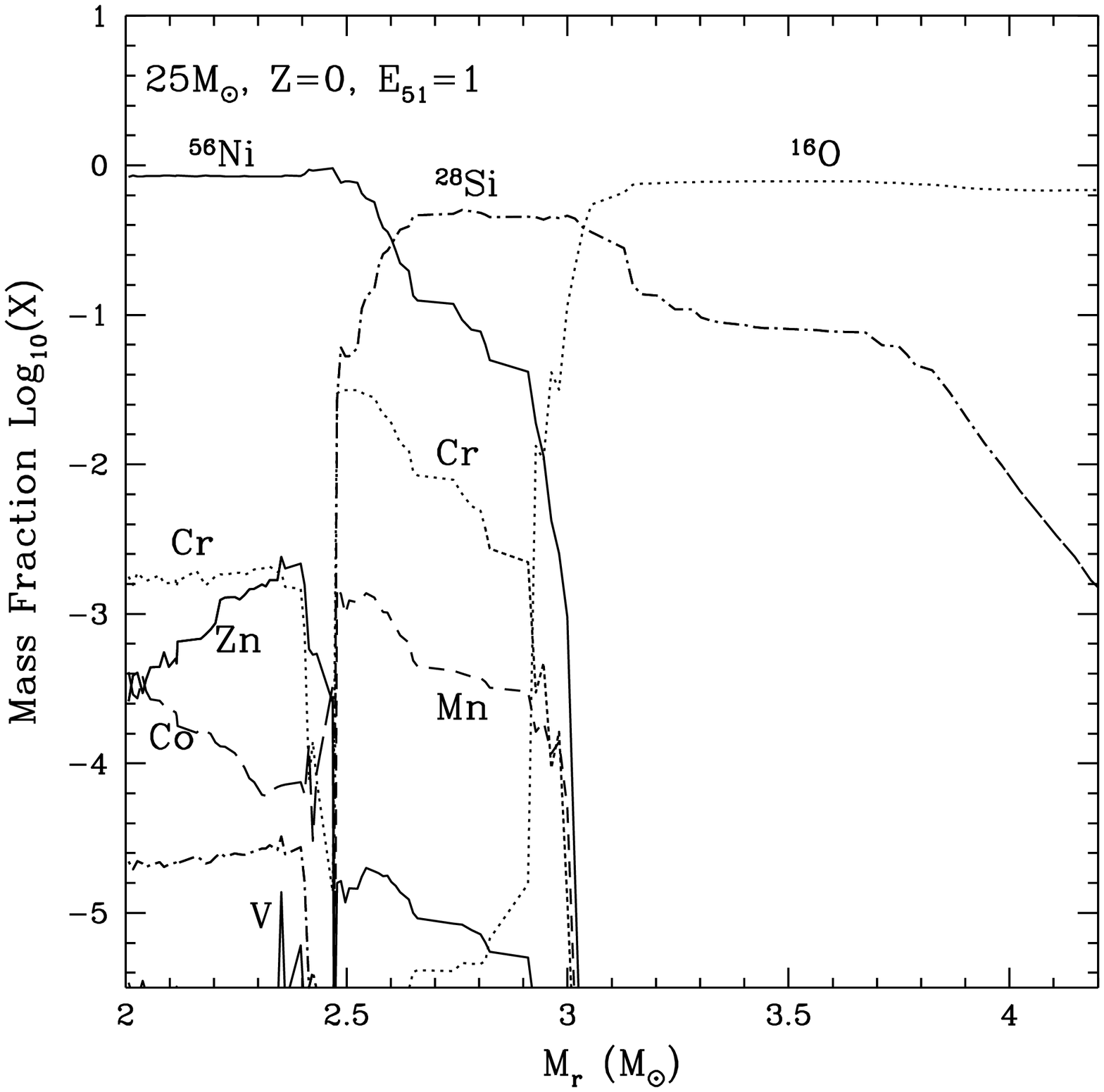}
\includegraphics*[width=6.5cm]{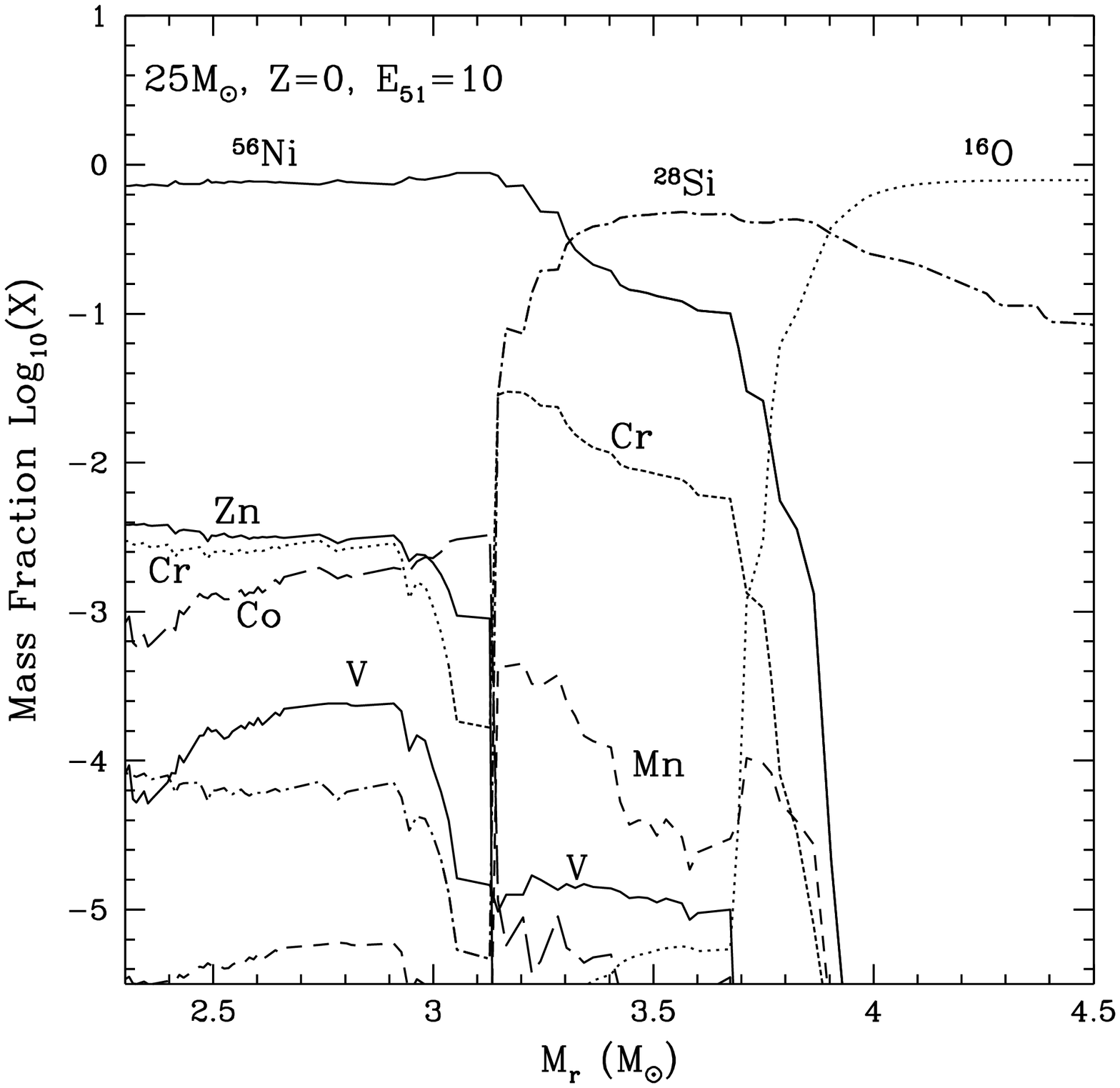}
\caption{Abundance distribution against the enclosed mass
$M_r$ after the explosion of Pop III 25 \ms\ stars with $E_{51} = 1$
(left) and $E_{51} = 10$ (right) \cite{umeda2002a}.}
\label{fig09}
\end{figure*}	

\subsection {Supernovae vs. Hypernovae}

The right panel of Figure~\ref{fig09} shows the composition in the
ejecta of a 25 $M_\odot$ hypernova model ($E_{51} = 10$).  The
nucleosynthesis in a normal 25 $M_\odot$ SN model ($E_{51} = 1$) is
also shown for comparison in the left panel of Figure~\ref{fig09}
\cite{umeda2002a}.

We note the following characteristics of nucleosynthesis with very
large explosion energies \cite{nakamura2001b,nomoto2001,ume05}:

(1) Both complete and incomplete Si-burning regions shift outward in
mass compared with normal supernovae, so that the mass ratio between
the complete and incomplete Si-burning regions becomes larger.  As a
result, higher energy explosions tend to produce larger [(Zn, Co,
V)/Fe] and smaller [(Mn, Cr)/Fe], which can explain the trend observed
in extremely metal-poor stars 
\cite{ume05,tominaga2005}.
(Here [A/B] $= \log_{10}(N_{\rm A}/N_{\rm B})-\log_{10} (N_{\rm
A}/N_{\rm B})_\odot$, where the subscript $\odot$ refers to the solar
value and $N_{\rm A}$ and $N_{\rm B}$ are the abundances of elements A
and B, respectively.)

(2) In the complete Si-burning region of hypernovae, elements produced
by $\alpha$-rich freezeout are enhanced.  Hence, isotopes synthesized
through capturing of $\alpha$-particles, such as $^{44}$Ti, $^{48}$Cr,
and $^{64}$Ge (decaying into $^{44}$Ca, $^{48}$Ti, and $^{64}$Zn,
respectively) are more abundant.

(3) Oxygen burning takes place in more extended regions for the larger
KE.  Then more O, C, Al are burned to produce a larger amount of
burning products such as Si, S, and Ar.  Therefore, hypernova
nucleosynthesis is characterized by large abundance ratios of
[Si,S/O], which can explain the abundance feature of M82 
\cite{umeda2002b}.

\begin{figure*}
\includegraphics*[width=12.5cm]{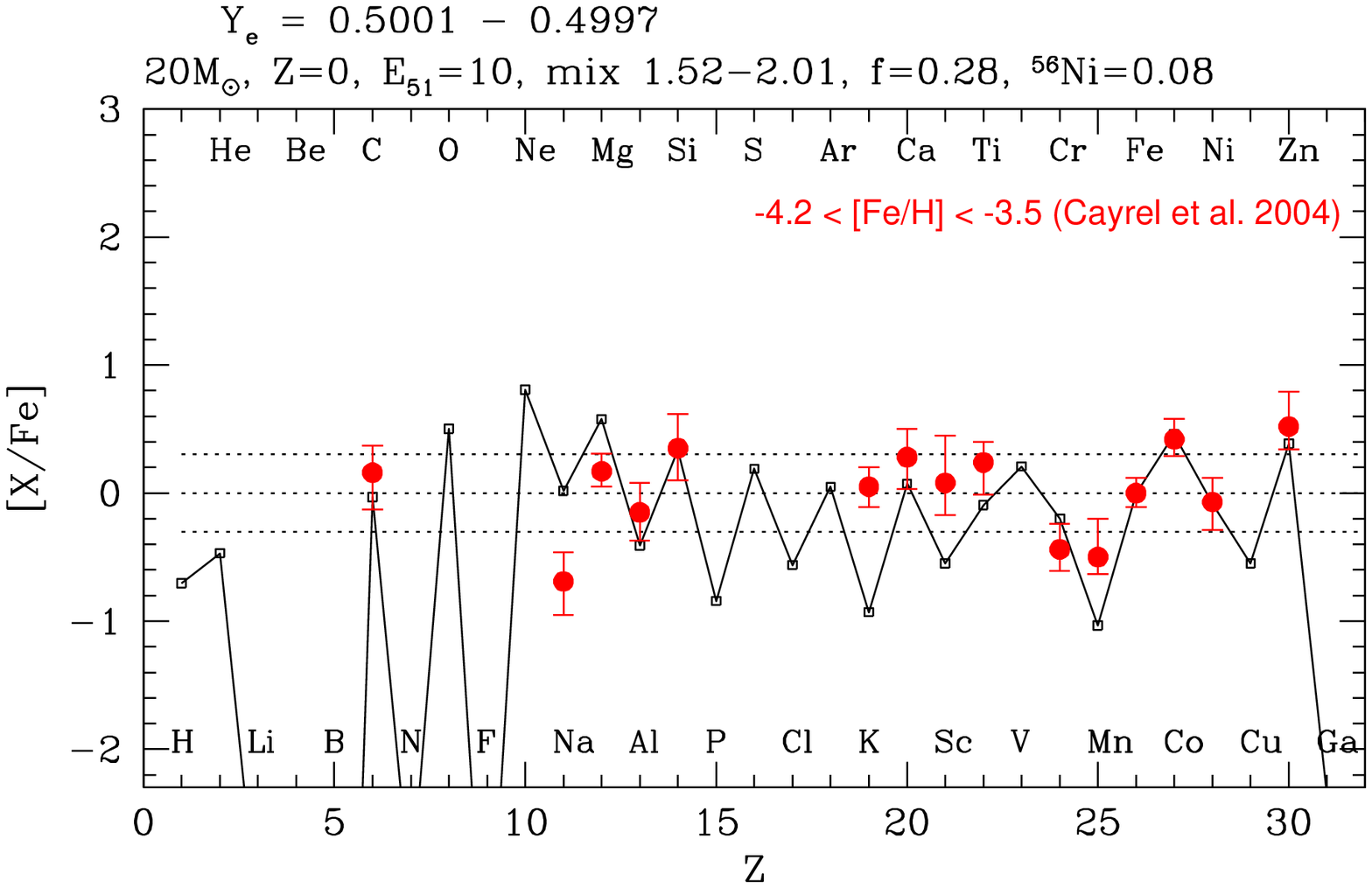}
\includegraphics*[width=12.5cm]{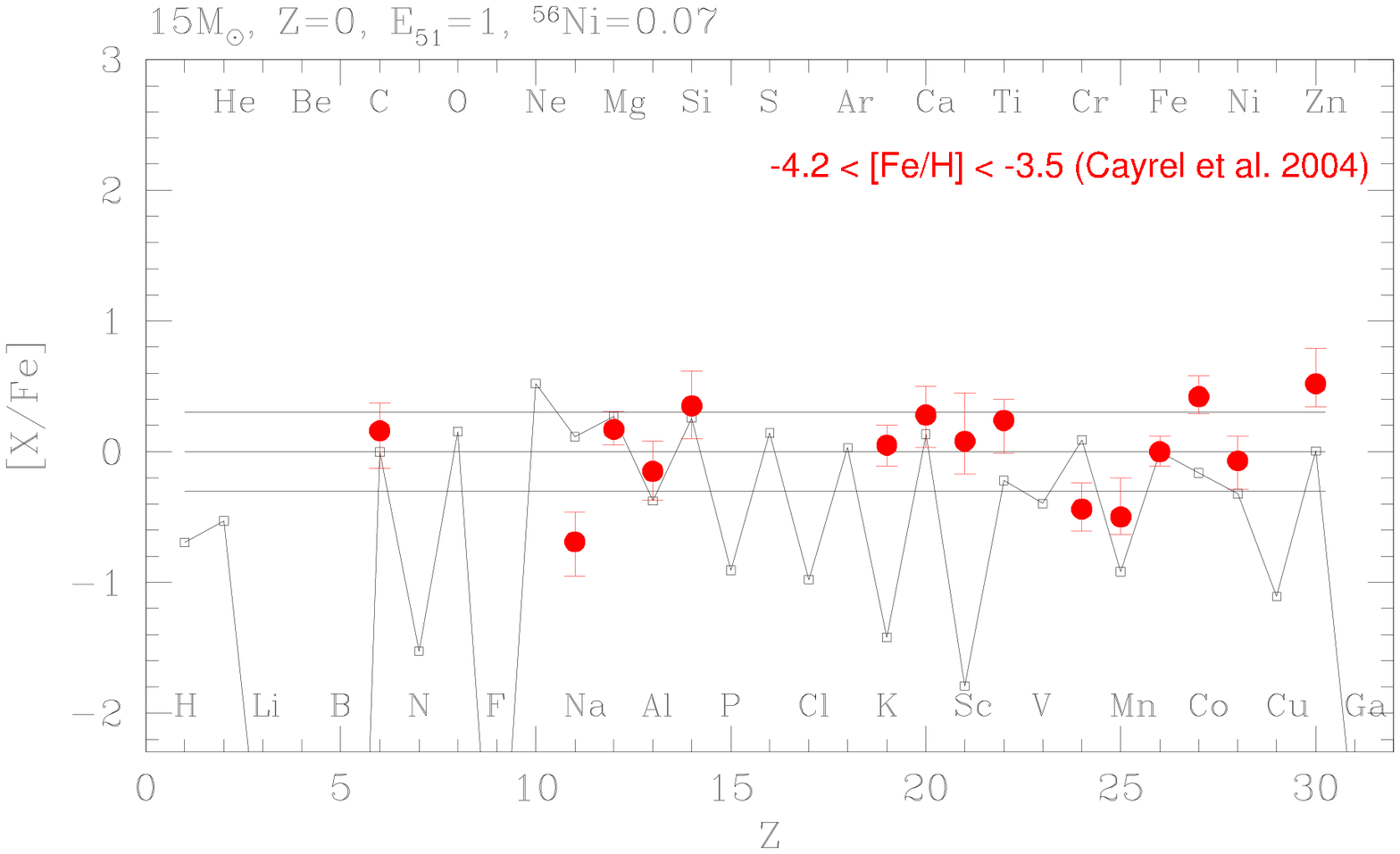}
\caption{Averaged elemental abundances of stars with [Fe/H] $= -3.7$
\cite{cayrel2004} compared with the hypernova yield (upper: 20 $M_\odot$,
$E_{51} =$ 10) and the normal SN yield (lower: 15
$M_\odot$, $E_{51} =$ 1).}
\label{fig11}
\end{figure*}

Hypernova nucleosynthesis may have made an important contribution to
Galactic chemical evolution.  In the early galactic epoch when the
galaxy was not yet chemically well-mixed, [Fe/H] may well be
determined by mostly a single SN event \cite{audouze1995}.  The
formation of metal-poor stars is supposed to be driven by a supernova
shock, so that [Fe/H] is determined by the ejected Fe mass and the
amount of circumstellar hydrogen swept-up by the shock wave
\cite{ryan1996}.  Then, hypernovae with larger $E$ are likely to
induce the formation of stars with smaller [Fe/H], because the mass of
interstellar hydrogen swept up by a hypernova is roughly proportional
to $E$ \cite{ryan1996,shigeyama1998} and the ratio of the ejected iron
mass to $E$ is smaller for hypernovae than for normal supernovae.

\begin{figure*}[!ht]
\centering
\includegraphics*[width=14cm]{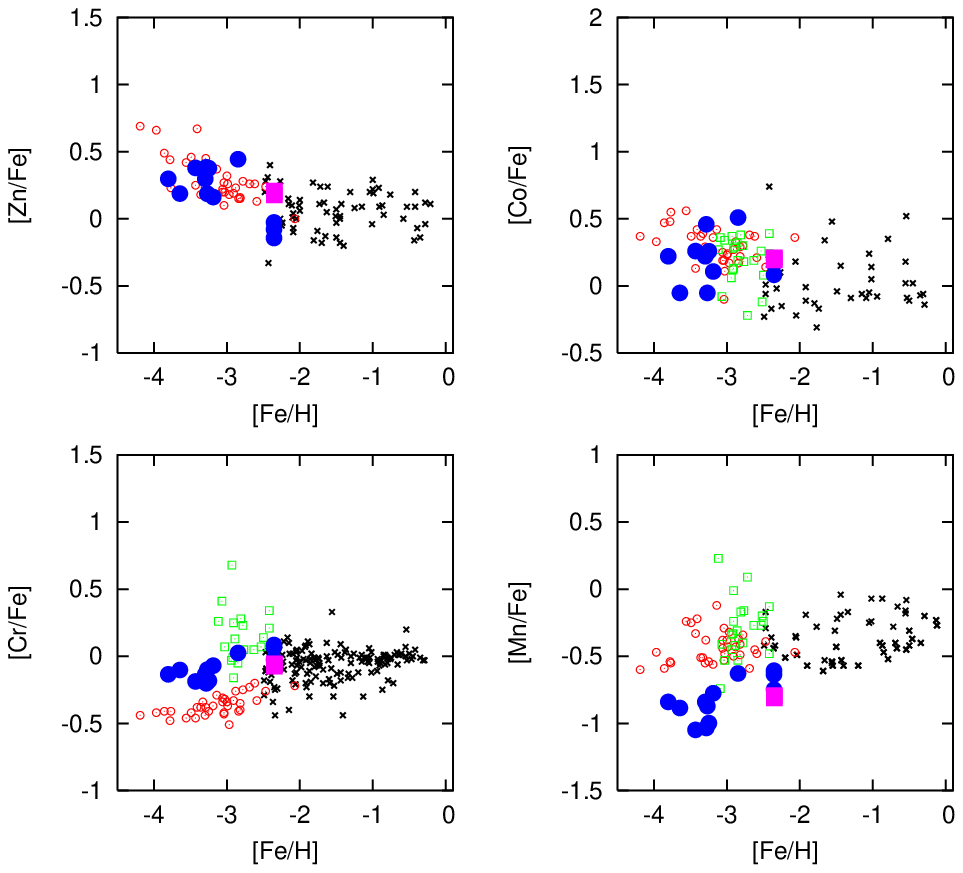}
\caption{Observed abundance ratios of [Zn, Co, Cr, Mn/Fe] vs [Fe/H]
  [{\it open circle}: \cite{cayrel2004}; {\it open square}: \cite{hon04}]
  compared with individual Pop III SN models ({\it filled circle}) and
  IMF-integrated models ({\it filled square}).}
\label{fig10}
\end{figure*}

\subsection{EMP Stars from VLT Observations}

The ``mixing and fall back'' process can reproduce the abundance
pattern of the typical EMP stars without a disagreement between
[(light element)/Fe] and [(Fe-peak element)/Fe].
Figure~\ref{fig11} shows that the averaged abundances of [Fe/H] $= -
3.7$ stars in \cite{cayrel2004} can be fitted well with the hypernova
model of 20 $M_\odot$ and $E_{51} =$ 10 (upper) but not with the normal
SN model of 15 $M_\odot$ and $E_{51} =$ 1 (lower) \cite{tominaga2005}.

\subsection{Hypernovae and Zn, Co, Mn, Cr}

In the observed abundances of halo stars, there are significant
differences between the abundance patterns in the iron-peak elements
below and above [Fe/H]$ \sim -2.5$ - $-3$ (Figure 12).

(1) For [Fe/H]$\lsim -2.5$, the mean values of [Cr/Fe] and [Mn/Fe]
decrease toward smaller metallicity, while [Co/Fe] increases
\cite{mcwilliam1995,ryan1996}.

(2) [Zn/Fe]$ \sim 0$ for [Fe/H] $\simeq -3$ to $0$ \cite{sneden1991},
while at [Fe/H] $< -3.3$, [Zn/Fe] increases toward smaller metallicity
\cite{cayrel2004}.

The larger [(Zn, Co)/Fe] and smaller [(Mn, Cr)/Fe] in the supernova
ejecta can be realized if the mass ratio between the complete Si
burning region and the incomplete Si burning region is larger, or
equivalently if deep material from the complete Si-burning region is
ejected by mixing or aspherical effects.  This can be realized if (1)
the mass cut between the ejecta and the compact remnant is located at
smaller $M_r$ \cite{nakamura1999}, (2) $E$ is larger to move the outer
edge of the complete Si burning region to larger $M_r$
\cite{nakamura2001b}, or (3) asphericity in the explosion is larger
\cite{mae03}.

Among these possibilities, a large explosion energy $E$ enhances
$\alpha$-rich freezeout, which results in an increase of the local
mass fractions of Zn and Co, while Cr and Mn are not enhanced
\cite{umeda2002a}.  Models with $E_{51} = 1 $ do not produce
sufficiently large [Zn/Fe].  To be compatible with the observations of
[Zn/Fe] $\sim 0.5$, the explosion energy must be much larger, i.e.,
$E_{51} \gsim 10$ for $M \gsim 20 M_\odot$, i.e., hypernova-like
explosions of massive stars ($M \gsim 20 M_\odot$) with $E_{51} > 10$
are responsible for the production of Zn.

Figure 12 exhibits that the higher-energy models tend to be
located at lower [Fe/H] = log$_{10}$ (Fe/$E_{51})-C'$ (if $C'$
distribute around a certain peak value), and thus can explain the
observed trend.

In the hypernova models, the overproduction of Fe, as found in the
simple ``deep'' mass-cut model, can be avoided with the mixing-fallback
model \cite{ume05}.
Therefore, if hypernovae made significant contributions to the early
Galactic chemical evolution, it could explain the large Zn and Co
abundances and the small Mn and Cr abundances observed in very
metal-poor stars (Fig.~\ref{fig11}:
\cite{tominaga2005}).

\subsection{Pair Instability SNe vs. Core Collapse SNe}

In contrast to the core-collapse supernovae of 20-130 $M_\odot$ stars,
the observed abundance patterns cannot be explained by the explosions
of more massive, 130 - 300 $M_\odot$ stars. These stars undergo
pair-instability supernovae (PISNe) and are disrupted completely
(e.g., \cite{umeda2002a,heger2002}), which cannot be consistent with
the large C/Fe observed in C-rich EMP stars.  The abundance ratios of
iron-peak elements ([Zn/Fe] $< -0.8$ and [Co/Fe] $< -0.2$) in the PISN
ejecta \cite{umeda2002a,heger2002}, cannot explain
the large Zn/Fe and Co/Fe in the typical EMP stars
\cite{mcwilliam1995,norris2001,cayrel2004} and CS22949-037 \cite{dep02} either.
Therefore the supernova progenitors that are responsible for the
formation of EMP stars are most likely in the range of $M \sim 20 -
130$ $M_\odot$, but not more massive than 130 $M_\odot$.  This upper
limit depends on the stability of massive stars.

\section{Summary and Discussion}

We summarize the properties of core-collapse SNe as a function of the
progenitor mass.  As seen in Figure 2, three GRB-SNe are all similar
Hypernovae (HNe) for their $\Mej$ and $E$.  For non-GRB HNe, whether
the non-dection of GRBs is the effect of different orientations or of
an intrinsic property is still a matter of debate, but there is a
tendency for them to have smaller $\Mej$ and $E$.

\subsection{XRFs and GRBs from 20 - 25 $\Msun$ Progenitors}

The discovery of XRF~060218/SN~2006aj and their properties extend the
GRB-HN connection to XRFs and to the HN progenitor mass as low as
$\sim 20 \Msun$.  The XRF~060218 may be driven by a neutron star
rather than a black hole.

The progenitor mass range of 20 - 25 $\Msun$ is particularly
interesting, because it corresponds to the transition from the NS
formation to the BH formation.  The NSs from this mass range could be
much more active than those from lower mass range because of possibly
much larger NS masses (near the maximum mass) or possibly large
magnetic field (i.e., Magnetar).  Possible XRFs and GRBs from this
mass range of 20 - 25 $\Msun$ might form a different population.

\subsection{The Rate of GRBs}

An estimate and a comparison for the rates of hypernovae and GRBs was
performed \cite{pod04}.  Within the substantial uncertainties, the
estimates are shown to be quite comparable and give a galactic rate of
$10^{-6}$ to $10^{-5} yr^{-1}$ for both events.  These rates are
several orders of magnitude lower than the rate of core-collapse
supernovae, suggesting that the evolution leading to an HN/GRB
requires special circumstances, very likely due to binary
interactions.

The discovery of SN~2006aj/XRF~060218 affects the estimate of the GRB
rates.  The observations of various bands suggest XRF~060218 is an
intrinsically weak and soft event, rather than a classical GRB
observed off-axis.  The existence of a population of less luminous
GRB/XRFs than ``classical'' GRBs is suggested.  Such events may be the
most abundant form of X- or gamma-ray explosive transient in the
Universe, but instrumental limits allow us to detect them only
locally, so that several intrinsically sub-luminous bursts may remain
undetected.  

If the low-redshift GRBs are really typical of the global GRB
population, then their discovery within the current time and sky
coverage must be consistent with the local GRB explosion rate as
deduced from the very large BATSE GRB sample.  Pian et
al.\cite{pian2006} include this underluminous population, assume no
correction for possible collimation, which may vary from object to
object, and obtain a local GRB rate of $110^{+180}_{-20}\;{\rm
Gpc^{-3} yr^{-1}}$, compared to $1\;{\rm Gpc^{-3} yr^{-1}}$ estimated
from the cosmological events only.  The local rate of events that give
rise to GRBs is therefore at least one hundred times the rate
estimated from the cosmological events only (i.e., those observed by
BATSE).  The fraction of supernovae that are associated with GRBs or
XRFs may be higher than currently thought.

Such an estimate is only sketchy and should be taken as an order of
magnitude estimate at present.  It should, however, improve as more
bursts with known redshifts are detected.

\subsection{Hypernova-First Star Connection}

Based on the results in the earlier section, we suggest that the first
generation supernovae were the explosion of $\sim$ 20-130 $M_\odot$
stars and some of them produced C-rich, Fe-poor ejecta.


This work has been supported in part by the Grant-in-Aid for
Scientific Research (17030005, 17033002, 18540231) and the 21st
Century COE Program (QUEST) from the JSPS and MEXT of Japan.

\end{document}